\documentclass[letter]{jpsj3}
\usepackage{txfonts}
\usepackage{here}
\usepackage[dvipdfm]{}
\usepackage{def}
\title{ Superconductivity in Tetragonal LaPt$_{2-x}$Ge$_{2+x}$ }
\author{
\name{Satoki \surname{Maeda}}$^1$,
\name{Kazuaki \surname{Matano}}$^1$,
\name{Hiroki \surname{Sawaoka}}$^1$,
\name{Yoshihiko \surname{Inada}}$^2$,
\name{Guo-qing \surname{Zheng}}$^{1,3}$
\thanks{E-mail:zheng@psun.phys.okayama-u.ac.jp}
}

\inst{
$^1$Department of Physics, Okayama University, Okayama 700-8530, Japan\\
$^2$Department of Science Education, Graduate School of Education, Okayama University, Okayama 700-8530, Japan\\
$^3$Institute of Physics and Beijing National Laboratory for Condensed Matter Physics,\\ 
Chinese Academy of Sciences, Beijing 100190, China\\

} 

\abst{We find that a tetragonal CaBe$_2$Ge$_2$-type  structure can be stabilized in non-stoichiometric LaPt$_{2-x}$Ge$_{2+x}$. 
We further discovered that the tetragonal LaPt$_{2-x}$Ge$_{2+x}$ with $x$=0.15 and 0.2 respectively superconduct  at \tc=1.85 K and 1.95 K, which is about four time higher than that in monoclinic LaPt$_2$Ge$_2$. }

\begin{document}
\maketitle
 
 Ternary silicides or germanides in the formula of MT$_2$X$_2$ (M=rare earth or alkaline earth metals, T=transition metals, X=Si or Ge) can crystallize in several different type structures.
 Most of  them crystallize in the body centered tetragonal ThCr$_2$Si$_2$ type structure, but some
with heavy T (T=Ir, Pt) crystallize in the primitive tetragonal CaBe$_2$Ge$_2$ type structure,  
 a variant of the ThCr$_2$Si$_2$ type.\cite{rare,lapt2si2,cept2ge2} 
 The ThCr$_2$Si$_2$ type structure consists of  M-TX-M-TX-M stacking planes, while  in 
 the CaBe$_2$Ge$_2$ type structure, the stacking sequence is  M-TX-M-XT-M.
%

 LaPt$_2$Ge$_2$ is a superconductor with a critical temperature  \tc=0.55 K and was thought to take  the CaBe$_2$Ge$_2$ type structure.\cite{LaPt2Ge2_discover} 
However, subsequent 
structure analysis 
shows that
 LaPt$_2$Ge$_2$  crystallizes  in a monoclinic type structure, with the  space group of either $P2_1$ \cite{LaPt2Ge2_structure} or $P2_1/c$.\cite{LaPt2Ge2_structure2}
%
Venturini $et$ $al$. reported that LaPt$_2$Ge$_2$ crystallizes in the non-centrosymmetric monoclinic structure with the space group $P2_1$,\cite{LaPt2Ge2_structure} 
a distorted version of the CaBe$_2$Ge$_2$ type. 
More recently, Imre $et$ $al$. proposed that their data for LaPt$_2$Ge$_2$ can be better fitted by  the centrosymmetric monoclinic structure with the space group  $P2_1/c$.\cite{LaPt2Ge2_structure2} 
This structure is also a distortion of the CaBe$_2$Ge$_2$ type  but the volume of the unit cell  is  doubled  compared to that  reported earlier by Venturini $et$ $al$. 

Meanwhile, it has been demonstrated that,
in the case of SmPt$_2$Ge$_2$,  one can switch between a monoclinic structure and a tetragonal structure by changing the stochiometry. 
Non-stoichiometric SmPt$_{2.25}$Ge$_{1.75}$ and SmPt$_{1.87}$Ge$_{2.13}$ crystallize in the tetragonal CaBe$_2$Ge$_2$ type structure, while stoichiometric 
SmPt$_2$Ge$_2$ crystallizes in the monoclinic structure with the space group $P2_1$.\cite{Sm_PdPt_SiGe}

In this work, we find that a tetragonal CaBe$_2$Ge$_2$-type  structure can also be stabilized in non-stoichiometric LaPt$_{2-x}$Ge$_{2+x}$. 
We further discovered that  LaPt$_{2-x}$Ge$_{2+x}$ with $x$=0.15 and 0.2 respectively superconduct  at \tc=1.85 K and 1.95 K, which is much higher than that in LaPt$_2$Ge$_2$ with a monoclinic structure. 
\begin{figure}[t]
\begin{center}\includegraphics[clip,width=65mm]{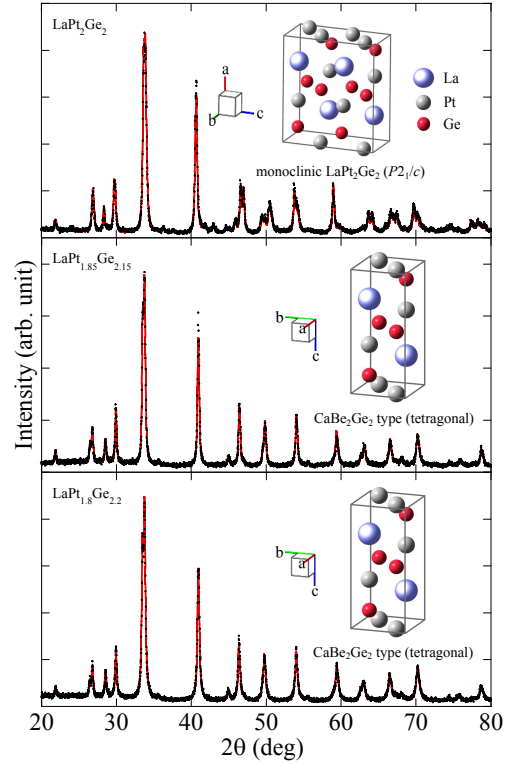}\end{center}
\caption{(color online) The observed powder X-ray diffraction patterns (Cu $K\alpha$ radiation) 
of LaPt$_2$Ge$_2$, LaPt$_{1.85}$Ge$_{2.15}$
and LaPt$_{1.8}$Ge$_{2.2}$. The solid curves show the calculated patterns for the structures shown in the insets.}
\label{f1}
\end{figure}
Stoichiometric and non-stoichiometric samples were synthesized by melting the elements of La (99.9\%), Pt (99.999\%) and Ge (99.999\%) 
in an arc furnace under high-purity (99.9999\%) Ar atmosphere. 
The resultant ingot  was turned over and re-melted several times to ensure good homogeneity.
 The weight loss during the arc melting was less than 1\%. 
 Subsequently, the samples were wrapped in Ta foil, sealed in a quartz tube filled with  He gas, 
  annealed at 1000 $^\circ$C for 3 days and then slowly cooled to room temperature over a period of 3 days. 
 The samples were characterized by powder X-ray diffraction (XRD)  using RIGAKU$\cdot$RINT-TTR III at room temperature. 
The XRD patterns were analyzed by RIETAN-FP program.\cite{RIETAN} The resistivity of LaPt$_{2-x}$Ge$_{2+x}$ was measured using a dc four-probe technique in the temperature range of 1.4 K-300 K.
For ac susceptibility measurements, a part of the ingot was powdered. The  ac susceptibility was determined by measuring the inductance of  a coil filled with a sample which is a typical setup for nuclear magnetic resonance measurements. 
Susceptibility measurement below 1.4 K was carried out with a $^3$He-$^4$He dilution refrigerator.
%
%
%
%
\begin{table}[H]
\caption{Space group, lattice parameters and $T_{\rm c}$ of the three compounds grown in this study. $\beta$ is the angle between the $a$- and $c$-axis.}
\label{t1}
\begin{center}
\begin{tabular}{lclclclcl}
\hline
\multicolumn{1}{l}{Compound} & \multicolumn{1}{c}{Space group} & Lattice parameters &$T_{\rm c}$ (K)\\
\hline
LaPt$_2$Ge$_2$ & $P2_1/c$ & $a$=10.037 \AA & 0.41 \\
&  & $b$=4.508 \AA &  \\
&  & $c$=8.987 \AA &  \\
&  & $\beta$=90.72 $^\circ$ &  \\ \hline
LaPt$_{1.85}$Ge$_{2.15}$ & $P4/nmm$ & $a$=4.396 \AA & 1.85 \\
&  & $c$=10.031 \AA &  \\ \hline
LaPt$_{1.8}$Ge$_{2.2}$ & $P4/nmm$ & $a$=4.395 \AA & 1.95 \\
&  & $c$=10.050 \AA &  \\ 
\hline
\end{tabular}
\end{center}
\end{table}

Figure \ref{f1} shows the XRD patterns of LaPt$_2$Ge$_2$, LaPt$_{1.85}$Ge$_{2.15}$ and LaPt$_{1.8}$Ge$_{2.2}$
The data for  LaPt$_2$Ge$_2$ can be well fitted by the monoclinic structure with space group $P2_1/c$. The peaks  unique to  $P2_1/c$ at   2$\theta$=41.9   and 43.0$^\circ$ were observed. 
The crystal structure of LaPt$_{1.85}$Ge$_{2.15}$ and LaPt$_{1.8}$Ge$_{2.2}$ was identified as the tetragonal CaBe$_2$Ge$_2$ type. Table \ref{t1} summarizes the crystal structure parameters of the three compounds.
%

\begin{figure}[H]
\begin{center}\includegraphics[clip,width=44mm]{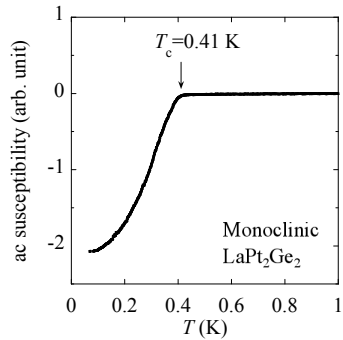}\end{center}
\caption{Temperature dependence of the ac susceptibility of LaPt$_2$Ge$_2$.}
\label{f2}
\end{figure}
\begin{figure}[H]
\begin{center}\includegraphics[clip,width=60mm]{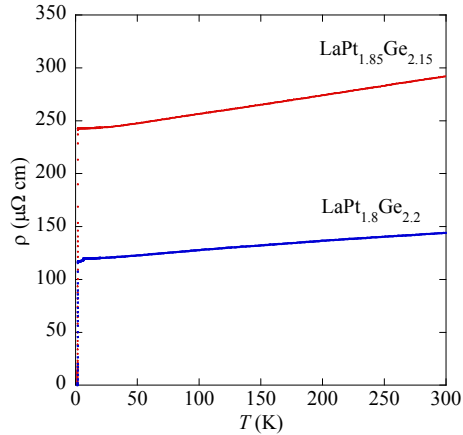}\end{center}
\caption{(color online) Temperature dependence of the electrical resistivity of LaPt$_{1.85}$Ge$_{2.15}$
and LaPt$_{1.8}$Ge$_{2.2}$.}
\label{f3}
\end{figure}
\begin{figure}[H]
\begin{center}\includegraphics[clip,width=60mm]{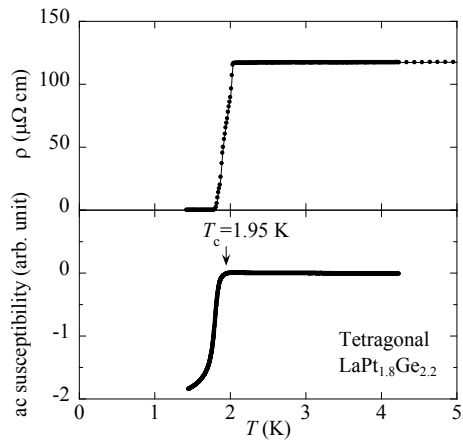}\end{center}
\caption{Temperature dependence of the resistivity and  the ac susceptibility of  LaPt$_{1.8}$Ge$_{2.2}$ below $T$=5 K. } 
\label{f4}
\end{figure}

Figure \ref{f2} shows the temperature dependence of the ac susceptibility of LaPt$_2$Ge$_2$.
Superconductivity was observed at $T_{\rm c}$=0.41 K, which is slightly lower than the previously reported value of 0.55 K\cite{LaPt2Ge2_discover}.
Figure \ref{f3} shows the electrical resistivity of LaPt$_{1.85}$Ge$_{2.15}$ and LaPt$_{1.8}$Ge$_{2.2}$ in the temperature range between 1.4 and 300 K. Zero resistivity was observed for  both compounds.
Figure \ref{f4} shows the enlarged part at low temperatures of the resistivity and  ac susceptibility  for LaPt$_{1.8}$Ge$_{2.2}$. 
The superconducting critical temperature determined as the onset of the diamagnetism seen in the susceptibility is  $T_{\rm c}$=1.85 K and 1.95 K for LaPt$_{1.85}$Ge$_{2.15}$ and LaPt$_{1.8}$Ge$_{2.2}$, respectively. 
The shielding fraction below $T_{\rm c}$ seen in the susceptibility is comparable to that for superconducting Indium powder measured under the same condition. An anomaly is also confirmed below $T_{\rm c}$ in the spin-lattice relaxation rate of LaPt$_{1.8}$Ge$_{2.2}$, which ensures the bulk nature of the superconductivity.

Finally, we briefly comment on possible cause of the  $T_{\rm c}$ enhancement in the tetragonal structure compared to that in the monoclinic structure. One possibility is that the tetragonal phase has a larger density of states. 
Another possibility is that the pairing state is different between the two cases. For example,  Li$_{2}$Pt$_{3}$B and Li$_{2}$Pd$_{3}$B have quite different $T_{\rm c}$, because the electronic band structure is quite different between them  and as a result a spin-triplet superconducting state in Li$_{2}$Pt$_{3}$B \cite{Nishiyama1} lowers its $T_{\rm c}$=2.7 K compared to the spin-singlet Li$_{2}$Pd$_{3}$B ($T_{\rm c}$=8 K) \cite{Nishiyama2}. 
More works, in particular band calculation, are needed to resolve the issue.

In summary, 
we succeeded in stabilizing a tetragonal phase of LaPt$_{2-x}$Ge$_{2+x}$ 
by changing the Pt:Ge ratio. 
Tetragonal LaPt$_{1.8}$Ge$_{2.2}$ exhibits superconductivity at \tc=1.95 K, which is higher than that for monoclinic  LaPt$_{2}$Ge$_{2}$  by more than three times.

\begin{acknowledgment}
We thank M. Nohara for useful discussion and S. Kawasaki for technical help. This work was partly  supported by KAKENHI Grant No. 22103004 and No. 20244058. 
\end{acknowledgment}

\end{document}